\documentclass[12pt,fleqn]{article}
\usepackage{amsmath,amsfonts}

\newcommand{\sla}[1]{\mathord{\not\mathrel{\kern .08333em #1}}} 
\newcommand{\M}{\mathcal{M}}
\newcommand{\R}{\mathbb{R}}
\newcommand{\Hi}{\mathcal{H}}
\newcommand{\ns}{\sla{n}}
\newcommand{\nsd}{\sla{\dot{n}}}
\newcommand{\Dels}{\mathord{\not\mathrel{\nabla}}}
\newcommand{\dirac}{\mathcal{D}}
\newcommand{\vol}{\epsilon}
\newcommand{\four}{{}^{\scriptscriptstyle \M}\!}  
\newcommand{\Vol}{\four\epsilon}
\newcommand{\inner}{\mathbin{\lrcorner}}
\newcommand{\Kbar}{\bar{K}}
\newcommand{\Khat}{\hat{K}}
\newcommand{\Rie}{\four{R}}
\newcommand{\Ries}{\four\mathord{\not\mathrel{R}}}
\DeclareMathOperator{\tr}{tr}
\DeclareMathOperator{\Tr}{Tr}
\DeclareMathOperator{\Res}{Res}
\newcommand{\NCint}{\mathop{\smash{\sim}\kern -.925em{\int}}}
\DeclareMathOperator{\End}{End}
\newcommand{\prince}{\sigma_{\text{pr}}}
\newcommand{\abs}[1]{\lvert #1 \rvert}
\newcommand{\smooth}{{\mathcal C}^\infty}
\newcommand{\A}{\mathcal{A}}
\newcommand{\g}{\mathfrak{g}}
\newcommand{\cc}{\mathcal{C}}
\DeclareMathOperator{\Op}{Op}
\newcommand{\LL}{\mathcal{L}^{(1,\infty)}}

\begin{document}
\thispagestyle{empty}
\begin{flushright}
{\small\sf CGPG-96/5-8}\\
{\small\sf gr-qc/9605068}\\
\end{flushright}
\bigskip
\begin{center}
{\huge
Hamiltonian Gravity and\\ Noncommutative Geometry}\\[3ex]
\bigskip
Eli Hawkins\\ 
\medskip
{\small\em Center for Gravitational Physics and Geometry}\\
{\small\em The Pennsylvania State University,
University Park, PA 16802}\\
{\small E-mail: mrmuon@phys.psu.edu}\\
\end{center}

\bigskip

\begin{abstract}
A version of foliated spacetime is constructed in which the spatial geometry is described as a time-dependent noncommutative geometry. The ADM version of the gravitational action is expressed in terms of these variables. It is shown that the vector constraint is obtained without the need for an extraneous shift vector in the action.
\end{abstract}

\section*{Introduction}
The problem of divergences in quantum field theory, and especially in quantum gravity, strongly suggests a need to describe the geometry of space-time as something different or more general than a classical manifold of points. One interesting generalization of classical geometry is noncommutative differential geometry. It has attracted some attention in physics in recent years, mostly in the form of Connes-Lott and related models of particle physics \cite{con3}. It has also been used as a regularization technique for Euclidean quantum field theory \cite{g-k-p}. Attention to applying noncommutative geometry to gravitation has been limited thus far; an expression for the Euclidean gravitational action was proposed by Connes and calculated explicitly by Kalau and Walze in \cite{k-w} and by Kastler in \cite{kas}.

In noncommutative differential geometry (see \cite{con2,con3,con5,con4}) the geometric information is set in the form of a {\em real spectral triple} $(\A,\Hi,D)$. This consists of a $*$-algebra (of ``functions'') $\A$ acting on a Hilbert space $\Hi$ and an unbounded, self-adjoint Dirac operator $D$ --- also acting on $\Hi$; the real structure is an antiunitary operator $J$, and for even dimension there is a $\mathbb{Z}_2$-grading operator on $\Hi$. 

An ordinary Riemannian spin geometry can be described by a spectral triple. In that case $\A$ is the commutative algebra $\smooth(\M)$ of smooth $\mathbb{C}$-valued functions with the involution (the $*$) being complex conjugation; the Hilbert space is the $\Hi=L^2(\M,S\M)$ of  spinor-functions with the canonical inner product $\langle\varphi\mid\psi\rangle = \int_\M \bar\varphi\psi\vol$; the Dirac operator is the usual $D=i\gamma^j\nabla_{\!j}$; the real structure $J$ is the charge conjugation operator; and the grading is the chiral grading --- the generalization of the familiar $\gamma_5$. For the details of the axioms a spectral triple must satisfy in general, see \cite{con4}.

If a real spectral triple is commutative (i.\ e., the algebra is commutative) and satisfies a slightly stronger axiom (that $JaJ^{-1}=a^*$), then up to equivalence it is the spectral triple of an ordinary manifold, although with some unwanted additional freedom to the Dirac operator (see section \ref{generalize}). In particular, the point set can be recovered as the spectrum of the algebra (as in Gelfand's theorem). In the noncommutative case, the spectral triple becomes a generalization of a Riemannian spin manifold.

We would like to use noncommutative geometry to generalise the geometry of space-time, but there is one rather obvious obstacle to this. Space-time at macroscopic scales resembles a Lorentzian rather than Riemannian manifold, and noncommutative geometry cannot straightforwardly describe a Lorentzian space-time. There is no natural {\em positive definite} inner product for  spinor-functions on space-time, and if an inner product is chosen anyway the Dirac operator cannot be self-adjoint since it has complex eigenvalues. 

There are many techniques of analysis which do not really work for space-times  --- enough that there is a standard way of makeshifting around the problem. 
The space-time is foliated into space-like hypersurfaces, and the leaf space is parameterised by a time variable. Since the leaves are spacelike, the geometric situation is reduced to Riemannian geometry.

The standard Hamiltonian approach to general relativity is the ADM (Arnowitt-Deser-Misner) formalism (see \cite{mis}). The {\em intrinsic} geometry of the leaves is simply described by a Riemannian metric. Two types of information describe how the leaves fit together. The {\em lapse function} gives the infinitesimal proper time separating the leaves at each point. 
The {\em shift vector} is needed because, rather than treating the leaves as separate topological spaces, the Hamiltonian approach treats them as a single manifold with time-dependent geometry. This means that the separate leaves must be identified pointwise, and the identified points trace out curves in space-time. The shift vector is the velocity of these curves, measuring their deviation from orthogonality to the foliation; it parameterises the arbitrary choice made in identifying points. By this construction, a time-dependent Riemannian metric, scalar field, and vector field on an $n$-dimensional manifold determine an $n+1$-dimensional space-time geometry.

The construction described in this paper is an algebraic Hamiltonian formalism --- a combination of noncommutative geometry and the standard Hamiltonian formalism. The time-dependent Riemannian geometry can obviously be described by a time-dependent, commutative spectral triple $(\A_t,\Hi_t,\dirac_t)$. The spacing of the leaves continues to be described by the lapse which is now thought of as taking values in $\A_t$ at each time. 
In the ADM formalism, the variables were time-dependent functions and fields, and it was necessary to fix the manifold on which these live. In this case the variables are operators and it is necessary to fix the Hilbert space on which these act. This means identifying all the $\Hi_t$'s to a single $\Hi$. One way to do this would be to extend the pointwise identification to identify the entire spinor bundles of the different leaves. This would be essentially a Schr\"{o}dinger representation, and would make the algebras $\A_t$ all identical. 

A much more covariant approach is to take the Heisenberg representation. The  spinor-functions at different times can be naturally identified by the 0-mass space-time Dirac equation. This identification involves no arbitrary choice and thus requires nothing like a shift vector.

In the ADM formalism the action is given by a Lagrangian which is equivalent to the Einstein-Hilbert Lagrangian modulo a classically irrelevant overall time derivative. The ADM Lagrangian can be split into kinetic and potential terms. The potential term is proportional to the integral of the spatial scalar curvature weighted by the lapse; an algebraic expression for this is easily adapted from Connes' expression for the Euclidean gravitational action. The kinetic term can be expressed in terms of the lapse and the extrinsic curvature tensor. The extrinsic curvature is simply related (via the lapse and shift) to the time derivative of the metric tensor, so in some sense it comes from the time derivative of the intrinsic geometry. In the algebraic formulation the intrinsic geometry is described by the Dirac operator; in order to extract the information of the extrinsic curvature which is needed to calculate the Lagrangian, we should look to the time derivative of the Dirac operator.

In section \ref{basic} most of the basic structures, notations, and formulas are introduced. In \ref{der} it is described how to calculate time derivatives of operators, and this is applied to the Dirac operator in section \ref{ddirac}. In \ref{spinors} the construction of the real structure and gradings are discussed as well as the issue of what type of spinors are used. The noncommutative analogue of integration is described in section \ref{residues}. In section \ref{scalarf} the action and dynamics of a scalar field are described in this geometrical setup; this is a simpler case of some of the issues that come up with the gravitational action which is calculated in section \ref{action}. The nature of diffeomorphism invariance is described in section \ref{diff1} and it is seen how the vector constraint comes about in this model. The remaining issues necessary to generalize this to truely noncommutative geometry are discussed in section \ref{generalize}. The failure of time diffeomorphism invariance in this model is discussed in section \ref{diff2}. The qualities and possible use of this model are discussed in section \ref{conclusions}. The appendix describes a simple calculation of the overall normalization of the Wodzicki residue and the normalization of the Euclidean gravitational action.

\section{Basic Definitions} \label{basic}
Let $\M=\R\times\Sigma$ be a space-time such that each of the hypersurfaces $\Sigma_t := \{t\}\times\Sigma$ is spacelike (in fact, Cauchy) and compact. These hypersurfaces will be regarded as the leaves of a foliation. Assume that $\M$ has a spin structure. The signature convention is time-like. Greek indices are space-time indices; Latin indices are spatial indices (``space''$=\Sigma_t$). When needed, an ``$\four\,$'' will distinguish a space-time object from a spatial object. $\nabla$ and $|$ will denote the space-time covariant derivative; $D$ and $;$ will denote the space covariant derivative. A slash denotes a tensor contracted with the Dirac vector $\gamma^\mu$, as in $\sla{a}=\gamma_\mu a^\mu$.

For any two  spinor-functions $\varphi,\psi \in \smooth(\Sigma_t,S\M)$  there is a positive definite inner product given by the flux of the current $\bar{\varphi}\: {*\gamma\psi}$; this generates a Hilbert space $\Hi_t=L^2(\Sigma_t,S\M)$ for each $t$. 
For solutions of the 0-mass Dirac equation $0=\Dels\psi$ this current is conserved, and so the inner product is independent of the specific $\Sigma_t$ at which the flux is taken. This means that solutions of the Dirac equation span another Hilbert space $\Hi$. For every $t$, there is an obvious map $\Hi\to\Hi_t$ given by restricting to $\Sigma_t$. This is actually an isomorphism; an arbitrary  spinor-function at $\Sigma_t$ is valid initial data for the Dirac equation, so propagation gives the inverse map $\Hi_t\to\Hi$. These isomorphisms naturally identify each $\Hi_t$ with $\Hi$. 

Each of the algebras $\A_t:=\smooth(\Sigma_t)$ has an obvious action on $\Hi_t$ and hence on $\Hi$. $\A_\R$ will be regarded as a bundle of algebras over $\R$. Since the $\Sigma_t$'s are all diffeomorphic, the algebras $\A_t$ are all (non-naturally) isomorphic.

The first calculations here are most conveniently done on Dirac spinors since some of the operators involved reverse chirality, but I will soon restrict to chiral spinors (section \ref{spinors}). For odd spatial dimension (hence even space-time dimension) there is a chiral grading $\Gamma$ of $\Hi$; this is the generalization of the usual $\gamma_5$. For all dimensions, there is an antiunitary charge conjugation operator $\cc$ acting on $\Hi$. Both $\Gamma$ and $\cc$ are time-independent, since they preserve solutions of the 0-mass Dirac equation.

The inner product can be written $\langle\varphi|\psi\rangle = \int_{\Sigma_t} \bar{\varphi}\:{*\gamma}\psi = \int_{\Sigma_t} \bar{\varphi} \ns \psi \vol$, where $n$ is the unit normal to $\Sigma_t$ and $\vol = n\inner\Vol$ is its volume element. This means, in particular, that for any Dirac matrix valued function $A$, the adjoint is $A^*=\ns\bar{A}\ns$; so, $\ns^*=\ns$ and $\gamma_j^* = -\gamma_j$. 

Let $\alpha^j$ be the Dirac vector of $\Sigma_t$. The spatial Dirac operator is $\dirac := i\alpha^jD_j = i\alpha^j(\partial_j + \dots)$. It should be the Hamiltonian associated to some sort of normal derivative; that is, the Dirac equation can be written in a Hamiltonian form as $i(\partial_n + \dots)\psi = \dirac\psi$. On the other hand $0=\Dels\psi = (\ns\partial_n + \gamma^j\partial_j + \dots)\psi$, so $i\partial_n\psi = (-i\ns\gamma^j \partial_j + \dots)\psi$; therefore, $\alpha^j = -\ns\gamma^j = \gamma^j\ns = (\alpha^j)^*$. This satisfies the required identity $[\alpha^j,\alpha^k]_+=2g^{jk}$ with the {\em positive definite} metric of $\Sigma_t$. Analogously to the slash notation, a tensor contracted with $\alpha$ will be denoted with a hat. Certain spatial tensors will also be used as space-time tensors, the identification is made in the ``natural'' index positioning of the tensor.

The $\nabla$ derivative of the normal vector is $n^\mu_{\:|\nu} = K^\mu_{\:\nu} + n_\nu a^\mu$. $K^\mu_{\:\nu}$ is the extrinsic curvature of the leaves (in natural position), and $a^\mu := n^\mu_{\; |n}$ is the acceleration of the unit normal (also in natural position). Both of these are really spatial tensors --- they vanish when contracted with $n$ --- and are written here in natural position. In covariant positon $K_{ij}$ is symmetric. Using a differential form notation, the lapse $N$ is the scalar defined by the equation $n = Ndt$; since $a$ is spatial, we have
\begin{align*}
a &= n\inner (n\wedge a) = n\inner dn = n\inner d(Ndt) 
= n\inner(dN\wedge dt) \\
&= -N^{-1}n\inner(n\wedge dN),
\end{align*}
i.\ e., the projection of $-N^{-1}dN$ orthogonal to $n$. Since $a$ is naturally contravariant rather than covariant, we have $a^j = N^{-1} N^{;j}$.

Parallel transport by $D$ respects the metric and keeps tangent vectors tangent to the foliation, therefore $D$ regards the unit normal vector as constant; so $0=[D_j,\ns]_-$. In general, $A\mapsto \frac12 \ns[\ns,A]_+$ is a projection onto operators commuting with $\ns$ (taking advantage of $\ns^2=1$), and $D_j$ (acting on spinors) is exactly the image of $\nabla_{\!j}$ under this projection:
\begin{align}
D_j &= \tfrac12 \ns [\nabla_{\!j},\ns]_+ = \nabla_{\!j} + \tfrac12 \ns[\nabla_{\!j},\ns]_- 
= \nabla_{\!j} + \tfrac12 \ns \gamma_\mu n^\mu_{\:|j} \nonumber\\
&= \nabla_{\!j} + \tfrac12 \Khat_j.  \label{Djform}
\end{align}
$\Khat_j := \alpha_k K^k_{\:j}$ is naturally covariant and self-adjoint. Consistency with the inner product means that $D_j$ must be skew-adjoint; in fact $-\frac12 \Khat_j$ and $D_j$ are precisely the self-adjoint and skew-adjoint parts of $\nabla_{\!j}$.

The spatial Dirac operator is 
\begin{align}
\dirac &= i\alpha^j D_j = i\alpha^j \left(\nabla_{\!j} + \tfrac12 \Khat_j\right) 
        = -i\ns \gamma^j \nabla_{\!j} + \tfrac{i}2 \Kbar \nonumber\\
       &= -i\ns \Dels + i\nabla_{\!n} + \tfrac{i}2 \Kbar \label{Dform}
\end{align}
(where $\Kbar = \tr K$). It is self-adjoint, anticommutes with $\ns$, and commutes with $\Gamma$.

\section{Time Derivatives} \label{der}
In order to find an expression for the gravitational action, we need to be able to extract the extrinsic curvature from the time derivative of $\dirac$. To do this we need an expression for the time derivative of $\dirac$ in terms of more familiar objects. For this we need a way of calculating time derivatives of operators. Toward this end, first consider  spinor-functions on $\M$.

A space-time  spinor-function $\psi\in\smooth(\M, S\M)$ restricts to a spatial  spinor-function $\Sigma_t\to S\M$ for every value of $t\in\R$; using the identification of $\Hi_t$ with $\Hi$, this defines a function $\R\to\Hi$ (i.\ e., $t\mapsto\psi(t)$) which is completely equivalent to $\psi$. One might ask what operator on $\smooth(\M, S\M)$ corresponds to $\frac{d}{dt}$ on $\smooth(\R,\Hi)$. If $\varphi\in\smooth(\M, S\M)$ is a solution of the Dirac equation (that is, $\varphi\in\Hi$), then the inner product with $\psi$ taken at $\Sigma_t$ is the inner product with $\psi(t)$, i.\ e., $\langle\varphi|\psi\rangle_t = \langle\varphi|\psi(t)\rangle$. We need to compute $\frac{d}{dt}\langle\varphi|\psi\rangle_t = \langle\varphi|\frac{d}{dt}\psi(t)\rangle$. To do this, use Stokes theorem on the infinitesimal volume between $\Sigma_t$ and $\Sigma_{t+dt}$, and the formula $\Vol=N dt\wedge\vol$,
\begin{align*}
\frac{d}{dt}\int_{\Sigma_t} \bar\varphi\:{*\gamma} \psi 
&= \int_{\Sigma_t} (\bar\varphi\gamma^\mu\psi)_{|\mu} \frac{\Vol}{dt} 
= \int_{\Sigma_t} \bar\varphi \left(\overleftarrow\Dels + \Dels\right)\psi\, N\vol \\
&= \int_{\Sigma_t} \bar\varphi N \ns\ns \Dels \psi\, \vol \\
&= \langle\varphi|N\ns\Dels\psi\rangle_t.
\end{align*}
So $\frac{d}{dt}=N\ns\Dels$.

An operator $A$ that acts on space-time spinors can be applied to a solution $\psi$ of the Dirac equation to give a space-time spinor $A\psi$. With the above identification $A\psi$ can be thought of as a time-dependent element of $\Hi$. This means that $A$ determines a time-dependent operator on $\Hi$. This construction gives a projection from space-time operators to time-dependent operators, and it has a preferred left inverse. Given a time-dependent operator on $\Hi$, there is a unique operator in its preimage that only involves spatial derivatives. If time-dependent operators are identified with space-time operators of this type, then time derivatives are easily calculated; the time derivative of an operator is $\dot{A}=[\frac{d}{dt},A]_-$.

In particular, suppose $f$ is a scalar function on $\M$. Then 
\begin{align}
\dot{f} = [N\ns\Dels,f]_-=N\ns\gamma^\mu f_{|\mu} = N f_{|n} - N\alpha^j f_{;j}
\label{scalar}
.\end{align}
Clearly, this is quite different from choosing a specific time direction and taking the partial derivative of $f$.

\section{The Time Derivative of $\dirac$} \label{ddirac}
Now we can calculate
\begin{align*}
\dot\dirac &= [N\ns\Dels,\dirac]_- 
= -[\dirac,N]_-\ns\Dels + N[\ns\Dels,\dirac]_- \\
&= -iN\hat{a}\ns\Dels + iN[\ns\Dels,\nabla_{\!n} + \tfrac12 \Kbar]_-
\end{align*}
using the fact $[\dirac,N]_-=iN\hat{a}$, and the formula \eqref{Dform} for $\dirac$. Now, $\hat{a}\ns=\sla{a}$ and the definition $a=[\nabla_{\!n},n]_-$ give
\begin{align*}
\dot\dirac &= -iN\sla{a}\Dels + iN\sla{a}\Dels + iN\ns[\Dels,\nabla_{\!n} + \tfrac12 \Kbar]_- \\
&= iN\ns[\Dels,\nabla_{\!n} + \tfrac12 \Kbar]_- .
\end{align*}
We can evaluate part of this using the identity that $[\nabla_{\!\mu},\nabla_{\!\nu}]_-=\frac14\Ries_{\nu\mu}:=\frac14 \gamma^\alpha \gamma^\beta \Rie_{\alpha\beta\mu\nu}$. This gives
\begin{align}
[\Dels,\nabla_{\!n}]_- 
&= [\Dels,n^\mu]_-\nabla_{\!\mu} + \gamma^\nu n^\mu [\nabla_{\!\nu},\nabla_{\!\mu}]_- \nonumber\\
&= [\Dels,n^\mu]_-\nabla_{\!\mu} + \tfrac14 \gamma^\nu n^\mu \, \Ries_{\nu\mu} \nonumber\\
&= [\Dels,n^\mu]_-\nabla_{\!\mu} + \tfrac12 \Ries_\mu n^\mu \label{inter}
\end{align}
since $\gamma^\nu \Ries_{\nu\mu} = \gamma^\nu\gamma^\alpha\gamma^\beta \Rie_{\alpha\beta\nu\mu} = \left(\four g^{\nu\alpha} -\four g^{\nu\beta} + \four g^{\alpha\beta} + \gamma^{[\nu} \gamma^\alpha \gamma^{\beta]}\right) \Rie_{\alpha\beta\nu\mu} = 2\:\Ries_\mu$ (because of the cyclic identity). Now use $\nabla\cdot n = \Kbar$ and the definition of $\Rie$ to get
\begin{align*}
\Ries_\mu n^\mu &= \gamma^\nu \, \Rie^\lambda_{\;\,\mu\lambda\nu} n^\mu 
= \gamma^\nu (n^\mu_{\:|\nu\mu} - n^\mu_{\:|\mu\nu}) \\
&= \left[\nabla_{\!\mu},[\Dels,n^\mu]_-\right]_- - [\Dels,\Kbar]_-
,\end{align*}
which can be inserted back into \eqref{inter} to give
\begin{align*}
[\Dels,\nabla_{\!n}]_- 
&= \tfrac12 \left[\nabla_{\!\mu},[\Dels,n^\mu]_-\right]_+ - \tfrac12 [\Dels,\Kbar]_-
,\end{align*}
and in turn gives
\begin{align*}
\dot\dirac
&= \tfrac{i}2 N\ns \left[\nabla_{\!\mu},[\Dels,n^\mu]_-\right]_+ 
.\end{align*}

We need to reexpress this in terms of $D$ rather than $\nabla$; the only difficulty is the action of $D$ on the vector index. Note that $[\Dels,n^\mu]_-= \ns(a^\mu + \Khat^\mu)$ is orthogonal to $n$ (hence spatial). If $v$ is a (scalar valued) spatial vector, then 
\begin{align*}
(N dt\wedge\vol) v^\mu_{\:|\mu} = \Vol \, v^\mu_{\:|\mu} &= d(v\inner\Vol)
= d\left(v\inner[N dt\wedge\vol]\right) \\
&= dt\wedge d(N v\inner\vol) 
= dt\wedge\vol \, (N v^j)_{;j}
\end{align*}
and so $[\nabla_{\!\mu},v^\mu]_-=v^\mu_{\:|\mu} = N^{-1} (Nv^j)_{;j}$. Since by \eqref{Djform} $v^\mu\nabla_{\!\mu} = v^j\nabla_{\!j} = v^j (D_j-\frac12 \Khat_j)$, this gives
$\nabla_{\!\mu} v^\mu = N^{-1} (D_j - \tfrac12\Khat_j) N v^j$ which continues to be valid when $v$ is  matrix-valued. 

This gives
\begin{align*}
\dot\dirac &= \tfrac{i}2 \ns [D_j - \tfrac12\Khat_j,N\ns(a^j-\Khat^j)]_+ \\*
           &= \tfrac{i}2 [D_j,N(a^j-\Khat^j)]_+ + \tfrac{i}4 N[\Khat_j,a^j-\Khat^j]_-
\end{align*}
since $\ns$ commutes with $D_j$ and anticommutes with $\Khat_j$. The last term vanishes to give
\begin{align*}
\dot\dirac &= \tfrac{i}2 [D_j,N^{;j}-N\Khat^j]_+
= \tfrac{i}2 [D_j,[D^j,N]_- - N\Khat^j]_+ \\*
&= \tfrac{i}2 [D^2,N]_- - \tfrac{i}2 [D_j, N\Khat^j]_+
\end{align*}
but $D^2= -\dirac^2 + \frac14 R$ and $R$ --- being a scalar --- commutes with $N$. This gives the final result:
\begin{align}
\dot\dirac &= -\tfrac{i}2 [\dirac^2,N]_- - \tfrac{i}2 [D_j,N\Khat^j]_+ 
=: \dot\dirac_+ + \dot\dirac_- . \label{Ddot}
\end{align}
These two terms, respectively, commute and anticommute with $\ns$. 

\section{Dirac and Chiral Spinors} \label{spinors}
The charge conjugation operator $\cc$ commutes with the covariant derivative operators, and anticommutes with the imaginary unit $i$. Depending on the dimension, $\cc$ sometimes commutes and sometimes anticommutes with $\gamma^\mu$ (see \cite{spinors}), but this sign cancels itself so that $\cc$ always commutes with $\alpha^j$. Together this means that $\cc$ anticommutes with $\dirac$. To construct a real spectral triple describing the spatial geometry, we need an antiunitary operator $J$ which should (among other conditions) anticommute with $\dirac$ for spatial dimension $n\equiv1 \mod 4$ and commute with it for other dimensions. Since $\ns$ anticommutes with $\dirac$, this condition is satisfied if we set $J=\cc$ for $n\equiv1 \mod 4$ and $J=\ns\cc$ otherwise. This choice also satisfies the other conditions for $J$, and leads to $J\Gamma=\Gamma J$.

If the space dimension is even, then the Dirac spinors on $\M$ have the same dimension as Dirac spinors on $\Sigma$. In fact the Dirac spinor bundle $S\M$ restricted to $\Sigma_t$ is equivalent to $S\Sigma_t$.

On the other hand, if the space dimension is odd, $S\M$ restricted to $\Sigma_t$ is equivalent to two copies of $S\Sigma_t$. However, chiral spinors on $\M$ restrict to a single copy of $S\Sigma_t$. In order to work with the Hilbert space of Dirac  spinor-functions on $\Sigma$, we must restrict $\Hi$ to be the $\pm1$ eigenspace of $\Gamma$ (left or right handed spinors). This can be done, since all operators of interest from now on are $\Gamma$-even (commute with $\Gamma$). 

If the space dimension is even then there is no $\Gamma$, but a grading operator is needed for an even dimensional spectral triple. If we temporarily work in an orthonormal frame such that the 0 direction is the normal direction, then the grading is (modulo a phase factor) $\alpha^1\dots\alpha^n \approx (\gamma^0)^n \gamma^1\dots\gamma^n$. For even dimension $n$, the product $\gamma^0\dots\gamma^n$ is just a phase; so the grading is $\approx(\gamma^0)^{n-1} = \gamma^0 = \ns$, which is precise (up to a sign) since it is Hermitian.

To summarise, we have now expressed the geometry of $\M$ as a time-dependent, real, commutative spectral triple $(\A_t,\Hi,\dirac_t)$, and a lapse function $N(t)\in\A_t$. If the dimension $n$ is even, then the (time-dependent) grading is $\ns$ and the real structure is $J=\ns\cc$ given by a time-independent antiunitary $\cc$. For odd dimension, if $n\equiv1\mod4$ then $J$ is time-independent, but if $n\equiv3\mod4$ then $J$ does depend on time.

\section{The Wodzicki Residue and Dixmier Trace} \label{residues}
For some vector bundle $V\to\Sigma$ over an $n$-dimensonal manifold, given a choice of coordinate system and trivialization, an arbitrary psuedodifferential operator $A$ can be written as a Fourier integral operator 
\[
\left(A\psi\right)(x)= (2\pi)^{-n} \int e^{-ip\cdot(x-y)} \sigma(A;x,p) \, \psi(y)\, d^ny\, d^np
\]
in terms of its (total) symbol $\sigma(A): T^*\Sigma \to \End(V)$. Define $\sigma_k(A)$ as the component of order $k$ in $p$. The highest order component is called the principal symbol, $\prince(A)=\sigma_{\text{Ord}(A)}(A)$. Expressing $A$ in terms of its symbol is equivalent to writing $A$ as a formal power series in $p_j:= i\partial_j$ with $p$'s always ordered to the right of $x$'s. Although the total symbol is strongly dependent upon the coordinates used to define it, the principal symbol is actually coordinate invariant; it also has the simple product property $\prince(AB)=\prince(A)\prince(B)$. This is an example of the pseudodifferential calculus; calculations on operators can be performed in terms of their symbols.

Given an arbitrary Riemannian metric on $\Sigma$, define $\langle\sigma\rangle(x)$ as the normalized trace of $\sigma$ averaged over the unit sphere in $T^*_x\Sigma$. Because of the invariance of the principal symbol, $\langle\prince(A)\rangle$ is coordinate invariant; $\langle\sigma_{-n}(A)\rangle\vol$ is also coordinate invariant and is independent of the metric. 

The normalized Wodzicki residue is defined as
\begin{equation*}
\NCint A = \int_\Sigma \langle\sigma_{-n}(A)\rangle\vol.
\end{equation*}
It is a trace (i.\ e., satisfies the cyclic identity) and is proportional to the Dixmier trace (see \cite{con2} and the appendix), which (for $\Sigma$ compact) can be defined using a fiducial Dirac operator $D$, or any operator such that $\prince \abs{D} = \lVert p\rVert$ (where $\abs{D}:=(D^* D)^\frac12=(D^2)^\frac12$ is the operator absolute value). Define the zeta function $\zeta_{D,A}(s)$ as the analytic continuation of $\Tr\:(\abs{D}^{-s}A)$. The (extended) Dixmier trace is $\Tr_{\omega}(A) = \Res_{s=0} \zeta_{D^n,A}(s)= \frac1{n}\Res_{s=0} \zeta_{D,A}(s)$ (see \cite{con5}). 

$D$ can still be used to define the Dixmier trace when $D$ is singular. In general, if $A$ is multiplied by a finite codimension projection, then $\zeta_{D,A}$ will only change by a holomorphic function, and $\Tr_\omega A$ will be unaffected. It is thus  possible to project out any finite dimensional eigenspace of $D$ when defining $\zeta_{D,A}$, so in particular the $0$-eigenspace of $D$ can be projected out. This should work provided that 0 has finite degeneracy in $D$, as it always does for a compact manifiold.

The ``limiting process'' $\omega$ depends upon the fiducial $D$, but if $A$ belongs to the ideal $\LL(\Hi)$ of operators of differential order $-n$, then the Dixmier trace is independent of the fiducial $D$ and for diagonalizable $A$ is (see \cite{con2})
\begin{equation} \label{dix}
\Tr_\omega(A) = \lim_{N\to\infty} \frac1{\ln N} \sum_{k=1}^N \lambda_k(A)
\end{equation}
where the $\lambda$'s are the eigenvalues arranged in order of descending $\abs\lambda$. The Dixmier trace vanishes on the ideal $\LL_0 \subset \LL$ of operators of lower order than $-n$. One of the axioms of noncommutative geometry requires that $\abs\dirac^{-n}\in\LL$.

Now suppose that $\Sigma$ is given a Riemannian spin structure, and that $V=S\Sigma$ the spinor bundle. The canonical metric and Dirac operator can be used as the fiducial ones; this means that $\abs{D}$ can be used to shift degrees. If we need the integral of $\langle\prince(A)\rangle\vol$, but $A$ is not of order $-n$, it is merely necessary to multiply $A$ by an appropriate power of $\abs{D}$. 
If $f\in\smooth(\Sigma)$ then $\sigma(f;x,p)=f(x)$, so
\begin{equation*}
\NCint f\abs{D}^{-n} = \int_\Sigma f\vol .
\end{equation*}
Thus a volume integral may be reexpressed in a purely algebraic way, and the Dixmier trace may be regarded as algebraic integration.

In order to evaluate Wodzicki residues, we need to evaluate expressions of the form $\langle p_i \dots p_l\rangle$. If there is an odd number of $p$'s then this will simply vanish. If there is an even number of $p$'s then symmetry implies 
$\langle p_i p_j \dots p_l p_m \rangle \propto g_{(ij} \dots g_{lm)}$. The proportionality can be determined by recursion from the fact that $p^2$ is equivalent to $1$ inside an $\langle\dots\rangle$. In particular
$\langle p_a p_b \rangle = \tfrac1n g_{ab}$ and
$\langle p_a p_b p_c p_d \rangle 
= \tfrac1{n(n+2)} \left(g_{ab}g_{cd} + g_{ac}g_{bd} + g_{ad}g_{bc}\right)$.

The power $\abs{D}^{2-n}$ is particularly interesting. From the setup of the calculation in terms of symbols, it is readily apparent that $\sigma_{-n} \left(\abs{D}^{2-n}\right)$ is second order in derivatives of the metric (i.\ e., linear in second derivatives, quadratic in first derivatives) and first order in derivatives of the gauge potential (when there is one); it also vanishes for flat space with Cartesian coordinates, trivial spin connection, and 0 gauge potential. $\langle\sigma_{-n} \left(\abs{D}^{2-n}\right)\rangle$ is coordinate invariant and gauge invariant, and therefore {\em can only be} proportional to the scalar curvature $R$. The precise relation is (see the appendix and \cite{k-w,kas})
\begin{equation} \label{Rformula}
\langle\sigma_{-n}(\abs{D}^{2-n})\rangle = -\tfrac{n-2}{24}R 
.\end{equation}
This makes it possible to reexpress integrals of the form $\int_\Sigma f R\vol$ algebraically.

\section{A Scalar Field} \label{scalarf}
It is possible to describe matter fields in this geometric context. The simplest case is, of course, a scalar field. If $\phi$ is a scalar field on $\M$, it can be thought of as a function $\phi(t)\in\A_t$. Its time derivative is $\dot\phi = [\frac{d}{dt},\phi]_- = N\ns\gamma^\mu\phi_{|\mu} = N\phi_{|n} - N\alpha^j \phi_{;j}$. On the other hand $[\dirac,\phi]_-=i\alpha^j\phi_{;j}$. Combining these gives the Lagrangian for a real scalar field
\begin{equation*}
L = \tfrac12 \NCint \left(\dot\phi N^{-1} \dot\phi + 2[\dirac,\phi]_-N[\dirac,\phi]_- - m^2 \phi^2\right) \abs\dirac^{-n}
.\end{equation*}

In the commutative case, there is some freedom to reorder factors in expressions of this kind. When an individual term inside a Wodzicki residue is in the ideal $\LL$, then reordering scalar factors will only change the expression by operators of smaller order; these are in $\LL_0$ and vanish when the Wodzicki residue is taken. Functions are always of order 0, and in the noncommutative case the commutator of two functions is also a function and is thus of the same order; the freedom to reorder is significantly reduced. This means that when a Lagrangian is generalized from the commutative case to the noncommutative case, there is a factor ordering ambiguity. The factor ordering here is chosen as the simplest which makes this Lagrangian manifestly real.

The canonical momentum conjugate to $\phi$ is $\pi_\phi := \frac{\partial L}{\partial\dot\phi}$, but what does this mean? It is a linear function of the possible variations of $\dot\phi$ for fixed $\phi$. Looking at the above formula for $\dot\phi$ it is clear that it can only vary by an element of $\A_t$, so $\pi_\phi\in\A_t^*$. It is in fact 
\[
\pi_\phi : u \mapsto \tfrac12 \NCint \left(N^{-1}\dot\phi \abs{\dirac}^{-n} + \abs{\dirac}^{-n} \dot\phi N^{-1} \right)u
.\]
Specifying the coefficient of $u$ in this integrand is not equivalent to giving $\pi_\phi$; the coefficient can, for instance, be changed by any operator in $\LL_0$, because this will not affect the Dixmier trace. This means that $\pi_\phi$ cannot easily be represented by a corresponding operator, and there is no simple way to write $\dot\phi$ in terms of $\phi$ and $\pi_\phi$. This makes a closed form expression for the Hamiltonian essentially impossible. This same problem occurs with the gravitational action.

\section{The Gravitational Action} \label{action}
The principal symbols of the terms of $\dot\dirac=\dot\dirac_+ + \dot\dirac_-$ are
\begin{align*}
\prince(\dot\dirac_+) &= N^{;j} p_j \\
\prince(\dot\dirac_-) &= -N\Khat^j p_j .
\end{align*}
$\dot\dirac_+$ can be expressed in terms of $\dirac$ and $N$; so, $\dirac$, $\dot\dirac$, and $N$ contain all the information of $K$. It should therefore be possible to express the (ADM) gravitational Lagrangian (see \cite{mis})
\begin{align*}
L &= \tfrac14 \int_{\Sigma_t} \Rie \frac\Vol{dt} - \tfrac{d}{dt}[\dots] \\
&= \tfrac14  \int_{\Sigma_t} N \left[\Kbar^2 - \tr K^2 - R\right]\vol
\end{align*}
in terms of these variables. (Units are such that $4\pi G = c = 1$.)

If we decompose the Lagrangian into ``kinetic'' and ``potential'' terms, then \eqref{Rformula} immediately gives the potential term ($n$ is again the dimension of $\Sigma$),
\begin{align*}
L_{\text{pot}} = -\tfrac14  \int_{\Sigma_t} N R \,\vol 
= \tfrac6{n-2}\NCint N \abs\dirac^{2-n} .
\end{align*}

The simplest expressions in $\dot\dirac_-$ and $\dirac$ yield the pieces needed to construct the kinetic term:
\begin{align*}
\langle\prince(\dot\dirac_-^2)\rangle = \langle N^2\Khat^j p_j \Khat^k p_k\rangle 
= \tfrac1n N^2 \langle\Khat^j \Khat_j\rangle = \tfrac1n N^2 \tr K^2 
\end{align*}
and
\begin{align*}
\langle\prince (\dirac \dot\dirac_-)^2 \rangle 
&= \langle(\hat{p}N\Khat^j p_j)^2\rangle \\
&= \tfrac1{n(n+2)} N^2\langle \alpha^j\Khat_j \alpha^k\Khat_k + \alpha^j\Khat_k \alpha^k\Khat_j + \alpha^j \Khat^k \alpha_j \Khat_k \rangle \\
&= \tfrac1{n(n+2)} N^2\langle 2\Kbar^2 + 2 \alpha^j \Khat^k K_{jk} - \alpha^j \Khat^k \Khat_k \alpha_j \rangle \\
&= \tfrac1{n(n+2)} N^2 \left[2\Kbar^2 + (2-n)\tr K^2\right]
.\end{align*}
Combining these gives
\begin{align*}
N^2(\Kbar^2-\tr K^2) = \tfrac{n(n+2)}2 \langle\prince(\dirac\dot\dirac_-)^2\rangle + \tfrac{n(n-4)}2 \langle\prince (\dot\dirac_-^2)\rangle
,\end{align*}
so
\begin{align*}
L_{\text{kin}} 
= \tfrac14\NCint N^{-1} \abs\dirac^{-n-4} \left[\tfrac{n(n+2)}2 (\dirac\dot\dirac_-)^2
    + \tfrac{n(n-4)}2 \dirac^2 \dot\dirac_-^2\right].
\end{align*}
However, we want this in terms of $\dot\dirac$ not $\dot\dirac_-$. If we replace $\dot\dirac_-$ by $\dot\dirac = \dot\dirac_- + \dot\dirac_+$ in the expression for $L_{\text{kin}}$ then this can be expanded into a term quadratic in $\dot\dirac_-$, a term mixed in $\dot\dirac_+$ and $\dot\dirac_-$, and a term quadratic in $\dot\dirac_+$. The term quadratic in $\dot\dirac_-$ is simply the expression for $L_{\text{kin}}$ that we started with. In the mixed term, the principal symbol of the integrand contains precisely one factor of $\alpha$, therefore its trace is 0 and the term vanishes. 
For the term in $\dot\dirac_+$ there is useful freedom to rearrange factors; since $\prince(\dot\dirac_+)$ is a scalar, we can commute $\dot\dirac_+$ past anything else in this Wodzicki residue; so the term is 
\begin{equation*}
\tfrac{n(n-2)}4 \NCint N^{-1} \dot\dirac_+ \abs\dirac^{-n-2} \dot\dirac_+ .
\end{equation*}

Combining these results gives
\begin{multline*}
L = \NCint N^{-1} \abs\dirac^{-n-4} \left[\tfrac{n(n+2)}8 (\dirac\dot\dirac)^2
    + \tfrac{n(n-4)}8 \dirac^2 \dot\dirac^2\right] \\
  - \tfrac{n(n-2)}4 \NCint N^{-1} \dot\dirac_+ \abs\dirac^{-n-2} \dot\dirac_+
  + \tfrac6{n-2}\NCint N \abs\dirac^{2-n} .
\end{multline*}
We still need to eliminate $\dot\dirac_+$ from the second to last term of this expression. Inserting the definition \eqref{Ddot} of $\dot\dirac_+$ into that term gives
\begin{align*}
-\tfrac{n(n-2)}4 \NCint N^{-1} \dot\dirac_+ \abs\dirac^{-n-2} \dot\dirac_+
 &= \tfrac{n(n-2)}{16} \NCint N^{-1} [\dirac^2,N]_- \abs\dirac^{-n-2} [\dirac^2,N]_- \\
 &= \tfrac{n(n-2)}{16} \NCint N^{-1} \dirac^2 N \abs\dirac^{-n-2} [\dirac^2,N]_-
 \\&\quad   - \tfrac{n(n-2)}{16} \NCint \abs\dirac^{-n} [\dirac^2,N]_-
\end{align*}
by expanding the first commutator. The last term vanishes because the principal symbol involved is an odd function of $p$. Expanding the remaining commutator gives directly
\[
\tfrac{n(n-2)}{16} \NCint N \abs\dirac^{2-n}
- \tfrac{n(n-2)}{16} \NCint N^{-1} \dirac^2 N \abs\dirac^{-n-2} N \dirac^2
.\]

The final result is
\begin{multline} \label{act1}
L = \NCint N^{-1} \abs\dirac^{-n-4} \left[\tfrac{n(n+2)}8 (\dirac\dot\dirac)^2
    + \tfrac{n(n-4)}8 \dirac^2 \dot\dirac^2\right]
 \\ + \left(\tfrac6{n-2} + \tfrac{n(n-2)}{16}\right) \NCint N \abs\dirac^{2-n}
    - \tfrac{n(n-2)}{16} \NCint N \dirac^2 N^{-1} \dirac^2 N \abs\dirac^{-n-2} .
\end{multline}

In the particular case of $n=3$ this becomes
\begin{multline*}
L = \NCint N^{-1} \abs\dirac^{-7} \left[\tfrac{15}8 (\dirac\dot\dirac)^2
    - \tfrac38 \dirac^2 \dot\dirac^2\right]
+ \tfrac{99}{16}\NCint N \abs\dirac^{-1}
  - \tfrac3{16} \NCint N \dirac^2 N^{-1} \dirac^2 N \abs\dirac^{-5} .
\end{multline*}

In the case of $n=2$, general relativity simplifies considerably. Unfortunately this formula for $L$ cannot be applied in that case since $n-2$ occurs in a denominator.

Instead of calculating these Wodzicki residues geometrically, we can calculate them algebraically via the Dixmier trace. This allows us to apply this action to the generalized case in which the spectral triple is no longer commutative.
However, the first two terms here are written asymmetrically and so are not manifestly real in the noncommutative case. Fortunately, there is still some freedom to reorder the factors. One of the axioms of noncommutative geometry states that for any $f\in\A$, the operators $\left[\abs\dirac,f\right]_-$ and $\left[\dirac,f\right]_-$ are bounded. This means that $\left[f,\abs\dirac^{-1}\right]_-=\abs\dirac^{-1} \left[\abs\dirac,f\right]_- \abs\dirac^{-1}$ is of order $-2$ --- lower order than $f\abs\dirac^{-1}$. Since the first two integrands of \eqref{act1} are in $\LL$, we can effectively commute $N^{-1}$ and $\abs\dirac^{-1}$ (or $\dirac$) within these expressions. Because of this, \eqref{act1} is real valued in general.

\section{Spatial Diffeomorphisms} \label{diff1}
One striking difference between this form of Hamiltonian gravity and the ADM form is the absence of a shift vector. In the ADM formulation, the shift is needed because it is necessary to identify points of successive leaves of the foliation; the shift parameterises the freedom to choose that identification. In this algebraic formulation there is no need to refer to points; instead,  spinor-functions are naturally identified on successive leaves using the Dirac equation. 

However, the shift plays another important role in the ADM formulation; variation of the action with respect to the shift gives the vector constraint --- the part of Einstein's equations concerning $G_{0j}$. Since there is no shift in the algebraic action, it is not readily apparent that the vector constraint will be among the equations of motion coming from the algebraic action. It is, however, inevitable that the vector constraint will be present since in the constraint algebra it is generated from the scalar constraint given by variation of the lapse. It will be explicitly shown in this section how the vector constraint really does come from the variation of this action.

The shift is a vector field, which is an infinitesimal diffeomorphism; so, in order to address this issue, it is necessary to first determine what a diffeomorphism is in the algebraic formulation. The group of diffeomorphisms of a surface $\Sigma$ is precisely the group of (outer) automorphisms of the algebra of functions $\smooth(\Sigma)$, but this group does not act naturally on the Hilbert space of spinors. Instead we need to use a group that does --- namely, the group of automorphisms of the spinor bundle which preserve the density valued inner product. This is precisely the group of unitary operators which preserve the algebra of functions, the real structure, and (for $n$ even) the grading. This group is essentially the semidirect product of diffeomorphisms with local unitary transformations of the spinor bundle. 

The generalization of a shift should live in the corresponding Lie algebra. In a small local region, the spinor bundle can be trivialized and spinor-functions look like half-densities valued in this bundle. A Lie algebra element looks like a Lie derivative plus an anti-Hermitian  matrix-valued function, but the Lie derivative of a half-density with respect to $\xi$ is $\Delta_\xi:=D_\xi+\frac12(D\cdot\xi)=\frac12 [\xi^k,D_k]_+$. The Lie algebra is 
\begin{multline*}
\g_t = \left\{\Xi\in\Op(\Hi)\mid \Xi^*=-\Xi,\; [\Xi,\A_t]_-\subseteq \A_t,\; \Xi\ns=\ns\Xi,\; \Xi J=J\Xi\right\} \\
\!\!\!\!\! = \left\{\Delta_\xi + X \mid \xi\in\smooth(\Sigma_t, T\Sigma_t),\; X=-X^\dagger\in\smooth(\Sigma_t,\End(S\Sigma_t)),\; \right. \\ 
\left. \vphantom{\Delta_\xi)X^\dagger}
JX=XJ,\; \ns X=X\ns\right\}
.\end{multline*}

In order to reintroduce degrees of freedom corresponding to a shift, drop the natural identification of $\Hi_t$ with $\Hi$ and instead let the identification be variable. This is equivalent to composing the natural identification with a unitary operator $U(t)$ from the above group; which is equivalent to replacing the fundamental variables in the action  with $U(t)\dirac(t)U(t)^{-1}$ and $U(t)N(t)U(t)^{-1}$. Since the Lagrangian \eqref{act1} is a Wodzicki residue of an expression in $\dirac$, $\dot\dirac$, and $N$, it is invariant under a simultaneous unitary transformation of these variables (treating $\dot\dirac$ as independent). The time-dependent unitary transformation of $\dirac$ and $N$ is thus equivalent to substituting $\dot\dirac \mapsto \dot\dirac + [U^{-1}\dot{U},\dirac]_-$ in the action.

Now, start from $U(t)=1$ (the natural identification) and infinitesimally vary $U$ by $\Xi$; this changes $U^{-1}\dot{U}$ by $\dot\Xi$. This is thus formally equivalent to changing $\dot\dirac$ by $[\dot\Xi,\dirac]_-$ in the action $\int_{t_0}^{t_1} L\,dt$. In order to extract the corresponding equations of motion, this should be expressed in terms of $\Xi$ not $\dot\Xi$; so, integrate by parts. This separates the change in the action into a time integral and the difference of a boundary term between the final and initial times. This boundary term is the change in the Lagrangian when $\dot\dirac$ is varied by $[\Xi,\dirac]_-$.

In particular, if $\Xi=\Delta_\xi$ then the corresponding variation in $\dot\dirac$ is
\begin{align*}
\delta(\dot\dirac) &= \left[\Delta_\xi,\dirac\right]_- 
= \tfrac{i}2 \alpha^j \left[[\xi^k,D_k]_+,D_j\right]_- \\
&= \tfrac{i}2 \alpha^j \left[[\xi^k,D_j]_-,D_k\right]_+ + \tfrac{i}2 \alpha^j \left[\xi^k,[D_j,D_k]_-\right]_+
.\end{align*} 
The last term simplifies,
\begin{align*}
\tfrac{i}2 \alpha^j \left[\xi^k,[D_j,D_k]_-\right]_+ 
&= i\alpha^j \xi^k [D_j,D_k]_- + \tfrac{i}2 \alpha^j \left[[D_j,D_k]_-,\xi^k\right]_- \\
&= \tfrac{i}4 \alpha^j \xi^k \hat{R}_{jk} + \tfrac{i}2 \alpha^j \xi^l R^{k}_{\;ljk} \\
&= 0
\end{align*}
and so $\left[\Delta_\xi,\dirac\right]_- = -\frac{i}2 \alpha^j \left[[D_j,\xi^k]_-,D_k\right]_+ = -\frac{i}2 \left[\alpha^j \xi^k_{\; ;j},D_k\right]_+$.
Comparing this with the formula \eqref{Ddot} for $\dot\dirac$, the effect on the Lagrangian is equivalent to varying $K^k_{\:j}$ by $N^{-1} \xi^k_{\: ;j}$. This gives the change in the Lagrangian as
\begin{equation*}
\delta L = \tfrac12 \int_{\Sigma_t} \left(\Kbar \xi^j_{\; ;j} - K^j_{\: k} \xi^k_{\; ;j}\right) \vol
= -\tfrac12 \int_{\Sigma_t} \xi^k \left(\Kbar_{;k} - K^j_{\:k;j}\right)\vol
\end{equation*}
which is exactly the vector constraint. Variation with $\Xi=X$ simply gives 0; this is in some sense an identically satisfied ``Gauss' law'' constraint.

The action given by \eqref{act1} doesn't actually know about the natural identification of $\Hi_t$ with $\Hi$. Indeed, an infinitesimal unitary transformation of $\dirac$ and $N$ (coming from $\g_t$) is among the valid variations of these fields anyway. This means that the vector constraint will be among the equations of motion coming from the algebraic action, provided the variation is understood in the following sense. Normally an action $\int_{t_0}^{t_1} L\,dt$ is required to be stationary with respect to variations with the arguments held fixed at the endpoints $\{t_0,t_1\}$ of integration; in this case the geometry at a given time is unchanged by a unitary transformation. If the variation is performed with the arguments at $t_0$ and $t_1$ fixed only up to unitary equivalence, then there will be boundary-term equations of motion in addition to the Euler-Lagrange equations. These give precisely the vector constraint. This does also work in a more conventional setting; if the shift is simply set to 0 in the ADM action, then the vector constraint can still be obtained by only fixing the initial and final geometries modulo diffeomorphism equivalence.

\section{The Noncommutative Generalization} \label{generalize}
If the variables occuring in the algebraic action are restricted to come from a classical commutative space-time, then by construction this is completely equivalent to general relativity. However, the algebraic Lagrangian \eqref{act1} can be applied equally well to a more general noncommutative geometry. In order to do this it is necessary to express the restrictions on the variables appearing in the action in a purely algebraic way, and to understand how the other structures of the spectral triple evolve with time. These other structures ($\A_t$, $J$, and sometimes $\ns$) enter into the action indirectly by restricting $\dirac$ and $N$. 

The axioms of noncommutative geometry are actually not strong enough that in the ordinary manifold case the Dirac operator will be restricted to the form $i\Dels$. We actually have the freedom to add a (somewhat restricted) matrix-valued function to the Dirac operator without violating the axioms. This change is, however, detected by the Euclidean gravitational action; the matrix makes an ultralocal, positive definite, quadratic contribution to the action. This means that the desired type of Dirac operator is distinguished within its equivalence class as minimising the Euclidean gravitational action. In the case at hand, the contribution to the gravitational Lagrangian is the same ultralocal term multiplied by $N$. This means that the ``correct'' Dirac operators will again minimise the gravitational action, provided that the lapse is stricly positive. The equations of motion given by the algebraic action will contain the desired restriction on the Dirac operator, in addition to the Einstein equations.

In order to understand how operators evolve in time, recall that this is the Heisenberg representation; operators should evolve by the equation $\dot{A} = i[H,A]_-$. The Hamiltonian is determined by looking back to the Schr\"{o}dinger representation. Suppose that a space-time  spinor-function $\psi$ is time-constant with respect to some identification of spinor bundles with 0 shift. Then $\nabla_{\!n} \psi = X \psi$ for some  matrix-valued function $X$ (independent of $\psi$). The time derivative in the Heisenberg sense is given by the Hamiltonian, 
\begin{align*}
H\psi &= -i\tfrac{d}{dt}\psi = -iN\ns\Dels\psi = -iN(\nabla_{\!n} - \alpha^j\nabla_{\!j})\psi \\
&= \left[-iN(X-\tfrac12\Kbar)+N\dirac\right]\psi
.\end{align*}
 This, together with the requirement that $H$ must be self-adjoint, gives $H = \frac12[N,\dirac]_+$. It is trivial to add in a generalized shift $\Xi(t)\in\g_t$; this gives the general Hamiltonian,
\begin{equation*}
H = \tfrac12\left[N,\dirac\right]_+ - i\Xi
.\end{equation*} 

The evolution of a specific $f\in\A_t$ depends upon the choice of $\Xi$, but since $[\Xi,a]_-\in\A_t$, the evolution of $\A_t$ as a whole is independent of $\Xi$. The evolution of $\A_t$ can be described axiomatically by the condition
\[
\dot{f}(t)\in \tfrac{i}2\left[\left[N,\dirac\right]_+,f\right]_-+\A_t \qquad
\forall f\in\smooth\left(\R,\A_\R\right)
.\]

The $\Xi$ is irrelevant to the evolution of the other structures as well; for even dimensions $[\Xi,\ns]_-=0$ implies that $\nsd = \frac{i}2\left[[N,\dirac]_+,\ns\right]_- = i [N,\dirac]_+\ns$. If $N$ is strictly positive, then this formula can be inverted and $\ns(t)$ actually determines $\dirac$. In the same way, for the $n\equiv3\mod4$ case $\dot{J} = i [N,\dirac]_+ J$, and $J(t)$ determines $\dirac$.

I have used the symbol $\smooth$ rather freely without really explaining what it means here. The most obvious definition of smoothness is the one associated with the norm topology: {\em A time-dependent operator is $\smooth$ if all its derivatives are bounded operators}. In the commutative case, the first derivative of a smooth section of $\A_\R$ is indeed bounded, but the second derivative is generically unbounded. This definition fails for $\ns(t)$ where the first derivative is unbounded and for $\dirac_t$ which is itself unbounded. It is therefore necessary to use a much weaker definition of smoothness. 

For any value of $t$ we can define the dense subspace $\Hi^\infty\subset\Hi$ as the common domain of all powers of $\dirac_t$. In the commutative case $\Hi^\infty$ is the space of smooth  spinor-functions at $\Sigma_t$, but since smooth initial data evolves smoothly under the Dirac equation this should be independent of $t$. By any reasonable definition, a smooth space-time operator acting on a smooth  spinor-function should give a smooth  spinor-function; this suggests the very weak definition for smoothness of a time-dependent operator: {\em $A\in\smooth\left(\R,\Op(\Hi)\right)$ if and only if $\langle\psi|A|\varphi\rangle\in\smooth(\R) \quad\forall \psi,\varphi\in\Hi^\infty $.} If $\dirac_t$ satisfies this condition then $\Hi^\infty$ will indeed be independent of the $t$ at which it is defined. This definition of smoothness is associated to a topology defined by the system of seminorms $\{A\mapsto\abs{\langle\psi|A|\psi\rangle} \mid \psi\in\Hi^\infty \}$. Since this is a subset of the seminorms which define the weak topology, this topology is even weaker than the weak topology.

$N$ is restricted  by the assumption that $\cc$ is time-independent. $\dot\cc=0$ implies that $\cc$ commutes with $iH$. Since $\cc$ always anticommutes with $i$ and $\dirac$, this means that $N$ must commute with $\cc$. Assuming that $N$ continues to be $\ns$-even, this gives the condition  $N=\cc N \cc^{-1} = J N J^{-1}$. It is also reasonable to assume that $N$ will be self-adjoint in general. In the commutative case the class of functions which commute with $J$ is the same as the class of functions which are self-adjoint, namely real valued functions. Because of this there are many choices for the range of $N$ which are equivalent in the commutative case. $N(t)$ can be taken as a self-adjoint element that commutes with $J$ in the algebra generated by $\A_t$ and $J\A_tJ^{-1}$, or in the sum $\A_t + J\A_tJ^{-1}$, or simply in $\A_t$. The last, most restrictive case implies that $N(t)\in\A_t\cap J\A_tJ^{-1} \subseteq \mathcal{Z}(\A_t)$ (the center of $\A_t$) since by the axioms $J\A_tJ^{-1}$ is required to commute with $\A_t$; this choice is actually supported by the special case of a Connes-Lott model.

In a Connes-Lott type model of particle physics (see \cite{con3}), the algebra is ``slightly noncommutative''. It is a finite dimensional matrix algebra over the algebra of functions on a commutative manifold. The gauge group is a subgroup of the group of unitary elements in the algebra. Most choices of $N$ would break this gauge symmetry, but if $N$ is chosen from $N(t)\in\A_t$ to satisfy $N=N^*=JNJ^{-1}$, then it will be proportional to the unit matrix --- the space-time geometry will be simply the product of a commutative space-time with an ``internal'' geometry. On the other hand, with a broader choice of $N$, there might be interesting models possible in which $N$ is responsible for some symmetry breaking.

In the commutative case the real structure does not allow the Dirac operator to be changed by a gauge potential. However, in the slightly noncommutative case a gauge potential can be added, and with this choice of $N$ it will not affect the gravitational action. This means that if the gravitational action is not  supplemented with a gauge action then the dynamics of the gauge potential will be indeterminate.

This example implies that in the general noncommutative case there will be perturbations to which the action is insensitive, but this is not necessarily a problem and the precise significance can only be determined by looking at specific examples.

\section{Time Diffeomorphisms} \label{diff2}
In the ADM formalism there are many possible choices of foliation for any given space-time, and since they are all equivalent descriptions it is possible to transform from one foliation to another. 

An infinitesimal transformation of the foliation is determined by a real function $\eta\in\smooth(\M)$ or equivalently $\eta(t)\in\A_t$. At each point of a time slice, $\eta$ is the infinitesimal coordinate time that the slice is deformed by there; $\eta N$ is the infinitesimal proper time. The argument to determine the effect of this transformation on a space-time  spinor-function $\psi$ is exactly the same as the argument in section \ref{der} for evaluating the time derivative, but with $N$ replaced by $\eta N$. This means that $\psi$ transforms by $\delta\psi(t)=\eta(t)\dot\psi(t)$. This determines the effect on any space-time operator $A$ to be $\delta A = [\eta\frac{d}{dt},A]_-$. If $f\in\smooth(\R,\A_\R)$ is a space-time function
then it should transform by $\delta f = [\eta\frac{d}{dt},f]_- = \eta\dot{f} + [\eta,f]_-\frac{d}{dt}$. In the commutative case the second term vanishes identically and this is fine, but in the noncommutative case the second term will not vanish in general. A space-time function should transform to a space-time function in the new foliation; it cannot involve any time derivative operators. The only transformations of the foliation which are acceptable in general are those for which $\eta(t)\in\mathcal{Z}(\A_t)$. This does, however, always include the reparameterisations of $t$ which do not change the foliation.

This accounts for the ugliness of the expression \eqref{act1} for the gravitational action. In the ADM formalism there is a simple choice of variables such that $N$ occurs only linearly in the action. As a result, the constraint equation that results from varying $N$ does not contain $N$ itself, and it is possible to solve the equations of motion for the other variables with an essentially arbitrary choice of $N$. The arbitrariness of $N$ corresponds directly to the arbitrariness of the foliation. In the general noncommutative case this is not so; $N$ is not arbitrary so it must satisfy nontrivial constraints and it is inevitable that $N$ occurs nonlinearly in the action.

If the most restrictive condition on $N$ is assumed, then the restriction on $\eta$ should not be suprising; indeed, $\eta$ should be restricted to the same form as $N$. If this restriction is chosen, then we should always be able to transform to $N=1$, at least for some finite time interval. This means that an arbitrary solution can still be obtained after setting $N=1$ in the equations of motion (which simplifies them considerably). Unfortunately, as in the ADM formalism, an $N=1$ description will generally only work for a finite time, since the time slices tend to develop cusps. 

\section{Conclusions} \label{conclusions}
This model has rather mixed qualities. It inherits several good qualities from noncommutative geometry. Diffeomorphism invariance takes on a much simpler and more linear form than in ordinary differential geometry. This formalism eliminates the arbitrary identification of points which is necessary in other forms of Hamiltonian gravity, and with it the shift vector. 

On the other hand, the expression for the Lagrangian \eqref{act1} is ugly. In the passage from the classical case to the noncommutative generalization there are many potentially arbitrary choices; a commutator could in principle be randomly inserted almost anywhere. However, this may not be a problem with only $\dirac$, $\dot\dirac$, and $N$ appearing in the action.  

This model may be most useful as a step towards a toy model of regularized quantum gravity. If a field theory can be constructed on a noncommutative geometry with finite dimensional Hilbert space and algebra, then the theory cannot be divergent when it is quantized (see \cite{g-k-p}). If the action can be written as the Dixmier trace of an operator in $\LL$, then this can be approximated (up to normalization) by the ordinary trace in the finite degree-of-freedom case; this is a direct result of equation \eqref{dix}. This is not the case for the gravitational action; the Dixmier trace, instead, picks out a logarithmically divergent part of a series which is faster than logarithmically divergent. 

The badly behaved part of the gravitational action is essentially just $\Tr_\omega (\abs\dirac^{2-n}N)$. This is calculated from the zeta function $\zeta_{\dirac,N}(s)$. Suppose that the zeta function has only simple poles at $s=n-2$ and $s=n$ and has no other singularities to the right of $s=n-2$; this is always the case for a commutative manifold. (The singularity set of the zeta function is called the ``dimension spectrum'' in \cite{con5}.) The residue of the pole at $s=n$ is proportional to $\bar{N}$, the space average of $N$. This means that $\zeta_{\dirac,(N-\bar{N})}(s) = \zeta_{\dirac,N}(s)-\bar{N}\zeta_\dirac(s)$ is regular to the right of $s=n-2$. 
This means that the trace of $\abs\dirac^{2-n}(N-\bar{N})$ is only logarithmically divergent, at least in some averaged sense. So, $\Tr_\omega \abs\dirac^{2-n}$ (which is simply the Euclidean gravitational action) is entirely responsible for the excessive divergence. 
If an appropriate approximation for the Euclidean gravitational action in the finite degree of freedom case can be found, then this probably could be easily extended to adapt \eqref{act1}. Although the Chamseddine-Connes spectral action \cite{c-c} is done in that spirit, that formula cannot be used directly in this case since it approximates a higher derivative gravity, rather than pure general relativity.

In summary, the geometry in this model is a generalization of space-time which consists of a smoothly time-dependent real spectral triple with a fixed Hilbert space, and a time-dependent ``lapse operator'' which comes from the algebra of functions. The real structure is time-independent for spatial dimensions $n\equiv1 \mod4$; for $n\equiv3\mod4$ it is time-dependent; and for even dimensions it is time-dependent, but the product of the grading and the real structure is time-independent. The time dependences of the grading and real structures are determined by the Dirac operator and lapse.

The gravitational action is given by a messy, second order (in $\frac{d}{dt}$) Lagrangian depending only on the Dirac operator, its time derivative, and the lapse. Although a first order Hamiltonian description exists in principle, the momentum conjugate to the Dirac operator is valued in a difficult to parameterise space of linear functions of operators. The gravitational Hamiltonian probably cannot be written in closed form.

In principle, in the cases when the grading or real structure are time-dependent, the Dirac operator is determined by this time derivative. This suggests that $\dirac$ could be eliminated in favor of $J$ as a fundamental variable. However, this leads to a very nasty Lagrangian, quadratic in $\ddot{J}$, which cannot be written in closed form. That is probably not the appropriate way to go since $J$ and $\ns$ do not appear directly in the original action. Instead, choose initial values for $J$ and $\ns$ and evolve them using $\frac12[N,\dirac]_+$.

In the commutative case it is indeed possible to reconstruct a description of the geometry in terms of the ADM variables. At each time $t$ the point set topology is recovered from $\A_t$, and the spatial metric can be reconstructed from $\dirac_t$. If we choose a specific generalized shift $\Xi$, then evolution by the corresponding Hamiltonian $H=\frac12[N,\dirac]_+ - i\Xi$ determines an identification between all the $\A_t$'s. Identifying the functions is equivalent to identifying points, so this makes the geometry into a time-dependent Riemannian geometry on a fixed manifold. The lapse remains the lapse; the shift vector, acting as a derivation of functions, is $a\mapsto[\Xi,a]_-$. 

\appendix
\section*{Appendix: Normalizations} 
The proportionality constants of $\Tr_\omega f\abs{D}^{-n} \propto \int f\vol$ and $\langle\sigma_{-n}\abs{D}^{2-n}\rangle \propto R$ can most easily be determined by looking at unit spheres. The sphere $S^n$ is an ${\rm SO}(n+1)$ symmetric space; its spinor bundle is inherited from $\R^{n+1}$ and is therefore trivial. As a representation of $\widetilde{\rm SO}(n+1)$, the sections of the spinor bundle are the tensor product $\mathbb{C}[S^n]\otimes S$ of the scalar functions $\mathbb{C}[S^n]$ with the ($2^{\lfloor\frac{n}2\rfloor}$-dimensional) spinor representation $S$; the scalar functions are the direct sum of all reduced symmetric powers of the fundamental representation. 

If we choose an origin $o\in S^n$, then there is a natural decomposition $\mathfrak{so}(n+1)=\mathfrak{so}(n)\oplus\mathfrak{k}$ where the $\mathfrak{so}(n)$ is the stabilizer of $o$, and $\mathfrak{k} \cong T_oS^n$ is its orthogonal complement. Choose a basis $\{J_A\}\subset\mathfrak{so}(n+1)$ of Hermitian generators that splits into $\{J_\alpha\}\subset\mathfrak{so}(n)$ and $\{J_k\}\subset\mathfrak{k}$. Since this is a unit sphere, the Dirac vector at $o$ is $\gamma_k=2\left(1\otimes J_k\right)$; this is fixed because $\gamma_{[j} \gamma_{k]}$ must give the spinor rotation matrices. A vector field $\xi\in\mathfrak{so}(n+1)$ acts as a Lie derivative on  spinor-functions, but the difference between a Lie derivative and a covariant derivative is something proportional to $\nabla\xi$. The $\xi\in\mathfrak{k}$ are precisely those $\xi\in \mathfrak{so}(n+1)$ such that $\nabla\xi$ vanishes at $o$; therefore, the Lie derivative action of $\mathfrak{k}$ at $o$ is the same as the covariant derivative action. The covariant derivative at $o$ is thus $\nabla_{\!k}=-iJ_k=-i\left(J_k\otimes1+1\otimes J_k\right) = -iJ_k\otimes1 - \frac{i}2 \gamma_k$. The Dirac operator at $o$ is $D=i\gamma^k\nabla_{\!k} = 2J_k\otimes J^k + \frac{n}2$, but since the action of $J_\alpha$ on scalars vanishes at $o$, this is equivalent to $D=2J_A\otimes J^A +\frac{n}2$ which is valid at all points.

Using standard formulae for the quadratic Casimir (see \cite{o-v}), this gives the spectrum of the Dirac operator as $\text{Spec}(D) \subseteq {\mathbb Z}+\frac{n}2$ with degeneracy 
\begin{align*}
d(\lambda)&=2^{\lfloor\frac{n}2\rfloor} \binom{\abs\lambda + \frac{n}2-1}{n-1}\\
&= \tfrac1{(n-1)!}2^{\lfloor\frac{n}2\rfloor}\left[\abs\lambda^{n-1} - \tfrac{n(n-1)(n-2)}{24} \abs\lambda^{n-3} + \dots\right]
.\end{align*}
 Note that the lowest eigenvalue of $D^2=\frac14R-\nabla^2$ is $\frac{n^2}4>\frac{n(n-1)}4 = \frac14 R$, as it should be. The zeta function is (the analytic continuation of) $\zeta_D(s) = \Tr\:\abs{D}^{-s} = \sum_\lambda \left[d(\lambda)+d(-\lambda)\right] \lambda^{-s} = \sum_\lambda 2d(\lambda) \lambda^{-s}$, where the sum is over positive integers (half odd integers) for even (odd) dimension. If the degeneracy is written $d(\lambda) = a_0\abs\lambda^{n-1} + a_1\abs\lambda^{n-3} + \dots$, then the zeta function can be written 
\begin{equation*}
\zeta_D(s) 
= \sum_\lambda 2\left[a_0 \lambda^{n-1-s} + a_1 \lambda^{n-3-s} + \dots\right]
= \sum_{k=0}^{\lfloor\frac{n-1}2\rfloor} 2a_k \zeta(s-n+2k+1)
\end{equation*}
where, for even dimension, $\zeta$ is the Riemann zeta function. For odd dimension, replace $\zeta(s)$ with $(2^s-1)\zeta(s)$; both of these functions  are meromorphic, with a single simple pole at $s=1$ of residue 1. This shows that $\zeta_D(s)$ is a meromorphic function with simple poles at $s=n, n-2, \ldots >0$; the residue at $s=n-2k$ is $2a_k$.

The residue at $n$ shows that $\Tr_\omega \abs{D}^{-n} = \frac1n \Res_{s=n} \zeta_D(s) = \frac2{n}a_0 = \frac2{n!} 2^{\lfloor\frac{n}2\rfloor}$. Comparing this with 
\begin{equation*}
\NCint \abs{D}^{-n} = \int_{S^n} \vol = \frac{2 \pi^\frac{n+1}2}{\left(\frac{n-1}2\right)!}
\end{equation*}
shows that (in general)
\begin{equation*}
\NCint A = \frac{(4\pi)^{\frac{n}2} (\frac{n}2)!}{2^{\lfloor\frac{n}2\rfloor}} \Tr_\omega A
.\end{equation*} 

Comparing the residues at $n-2$ and $n$ shows
\begin{equation*}
\NCint \abs{D}^{2-n} = \frac{a_1}{a_0} \NCint \abs{D}^{-n} = -\tfrac{n(n-1)(n-2)}{24} \int_{S^n} \vol = -\tfrac{n-2}{24} \int_{S^n} R\vol
.\end{equation*}
This means that (in general)
\begin{equation*}
\int fR\vol = -\tfrac{24}{n-2} \NCint f\abs{D}^{2-n}
.\end{equation*}

\paragraph*{Acknowledgements:}
I wish to thank Lee Smolin for extensive discussions, and Abhay Ashtekar, Alain Connes, Murat Gunaydin, and Boris Tsygan for advice with various issues that arose. This material is based upon work supported under a National Science Foundation Graduate Fellowship. Also supported in part by NSF grant PHY95-14240 and by the Eberly Research Fund of the Pennsylvania State University.

\end{document}